\documentclass[aps,prl,reprint,superscriptaddress]{revtex4-1}

\usepackage{graphicx}
\usepackage{dcolumn}
\usepackage{mathtools, nccmath}
\usepackage{bm}
\usepackage{xcolor}
\usepackage{amsmath}
\usepackage{braket}
\usepackage[export]{adjustbox}
\usepackage{framed}
\usepackage[]{graphicx}
\usepackage[caption=false]{subfig}

\usepackage{hyperref}
\usepackage[mathlines]{lineno}
\usepackage[normalem]{ulem}
\usepackage{textcomp}
\usepackage{gensymb}
\usepackage[super]{nth}
\usepackage{graphicx,color}
\usepackage{hyperref}
\usepackage{babel}
\usepackage{booktabs}
\usepackage{dsfont}
\usepackage[percent]{overpic}
\usepackage{ulem}

\newcommand{\pdag}{{\phantom{\dagger}}}

\usepackage{stackengine}

\begin{document}

\title{Chiral damping with persistent edge states: interplay of spectral topology and band topology in open quantum systems}

\author{Ronika Sarkar$^*$}
\affiliation{Department of Physics, Indian Institute of Science, Bengaluru, 560012, India}
\affiliation{Solid State and Structural Chemistry Unit, Indian Institute of Science, Bengaluru 560012, India}
\author{Suraj S. Hegde$^*$}
\affiliation{Indian Institute of Science Education and Research Thiruvananthapuram, Vithura, 695551,
India}
\author{Awadhesh Narayan}
\affiliation{Solid State and Structural Chemistry Unit, Indian Institute of Science, Bengaluru 560012, India}
\author{Tobias Meng}
\affiliation{Institute of Theoretical Physics and W\"urzburg-Dresden Cluster of Excellence ct.qmat, Technische Universit\"at Dresden, 01062 Dresden, Germany}

\begin{abstract}
We study the dynamical consequences of combining the non-Hermitian skin effect with topological edge states. Focusing on the paradigmatic dissipative Hofstadter model, we find that the time-dependent particle density exhibits both chiral damping (due to the non-Hermitian skin effect) and edge-selective extremal damping (rooted in persistent topological edge states). We find that the time scales of chiral damping and edge-selective extremal damping decouple due to boundary-induced spectral topology, thus allowing observation of both effects under dynamics. We identify intermediate magnetic fields as the most favorable regime, since chiral damping is then partially recovered. More generally, our work sheds light on how open quantum systems are impacted by the combined presence of spectral and band topologies, and how their interplay can be probed directly.

\end{abstract}

\maketitle

{\it Introduction.} The topological characterization of matter has gained relevance beyond quantum materials, and is now also applied in photonic~\cite{xie2018photonics,ozawa2019topological}, mechanical~\cite{ma2019topological,lo2021topology}, soft matter~\cite{shankar2022topological}, and dissipative systems~\cite{akin1993general,bardyn2013topology,gneiting2022unraveling}. Often, the salient physics can at least approximately be described by non-Hermitian matrices. The interplay of such non-Hermitian ``Hamiltonians''  with topology is therefore under intense investigation. As a prominent example, the non-Hermitian skin effect (NHSE) is rooted in a topological winding of the complex spectrum of non-Hermitian matrices (``spectral topology''), rather than the topology of eigenstates (``band topology'')~\cite{Yao_2018,Zhang_2020,Bergholtz_2021,lin2023topological}. In open quantum systems, the NHSE of damping generators manifests as a chiral damping wavefront in the dynamics~\cite{Song2019,Yang2022}. 

In general, the spectral and band topologies of effective non-Hermitian systems can both be trivial or non-trivial. Non-Hermitian Hamiltonians can for example be associated with finite Chern numbers and have topological edge states~\cite{Yao_Song_Wang_2018}. In open quantum systems, this can lead to edge-selective extremal damping, a phenomenon in which particles resist dissipation predominantly at one or several edges of the system~\cite{Hegde23,Cherifi_2023,Yang_2023}. Some aspects of the interplay of NHSE and edge modes have been studied earlier. It was found that these effects compete, which can dramatically alter the localization of topological edge states~\cite{Zhu2021,Cheng2022,Zhu2022,Wang2022Morph}. 

\begin{figure}[t!]
    \centering
    \includegraphics[width=0.45\textwidth]{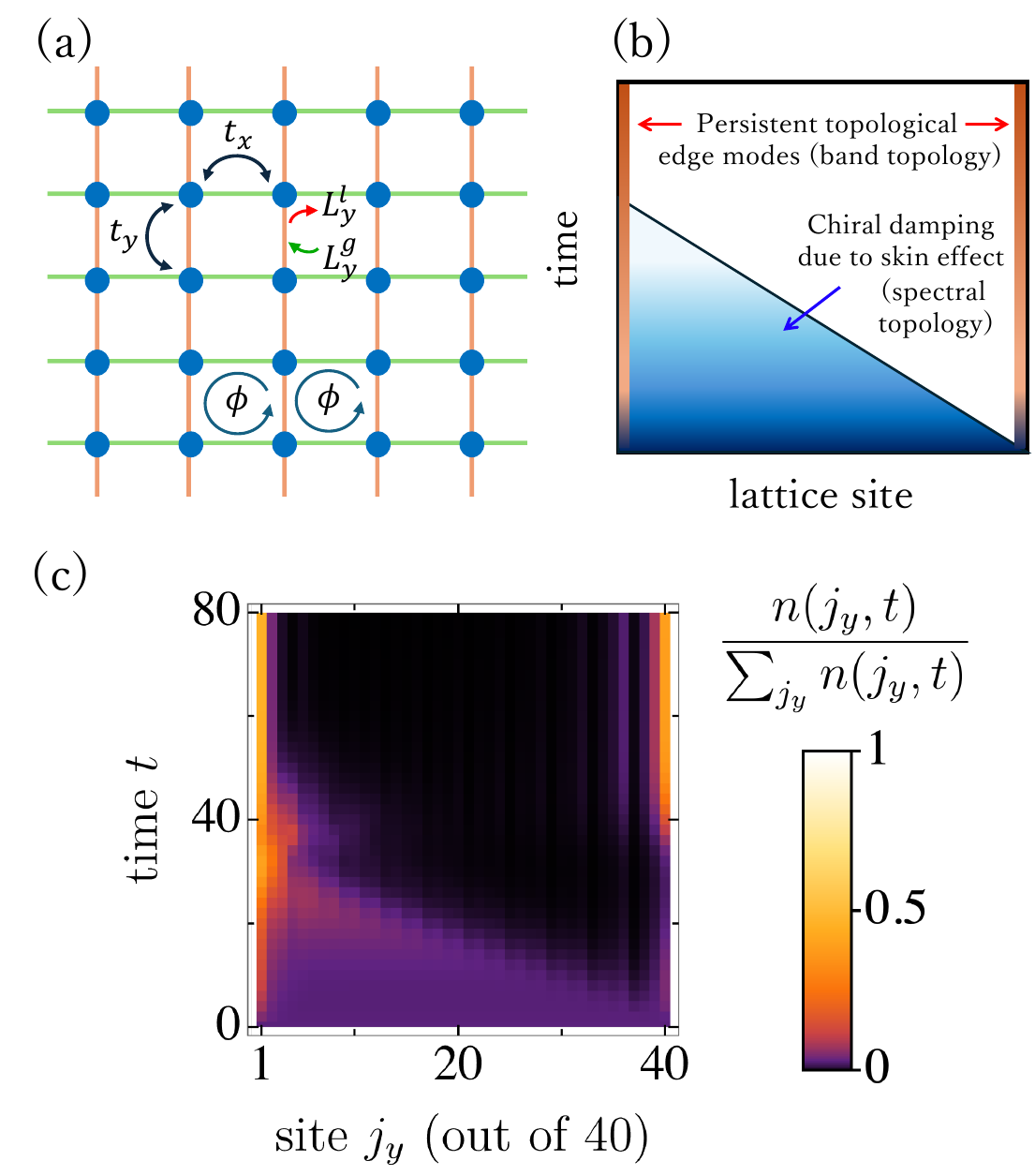}
    \caption{\label{fig:model} \textbf{System and main results.} (a) Dissipative Hofstadter model with hoppings $t_x$ and $t_y$, dissipators $L^l_y$ (loss) and $L^g_y$ (gain) on $y$-bonds, and enclosed magnetic fluxes $\phi$ per plaquette. (b) Schematic illustration of the coexistence of skin effect and topological edge modes. (c) Time evolution of the particle number $n(j_y,t)$ in the tight-binding model. We used periodic boundary conditions along $x$ (summing 40 $k_x$-momenta per site), open boundary conditions along $y$ (with $L_y=40$ sites and site index $j_y$). The system is fully occupied at $t=0$, and $\gamma^l=0.2$, $\gamma^g=0$, $\phi/\phi_0=1/4$, and $t_x=t_y=1$.}
\end{figure}

The combined dynamical consequences of the NHSE and topological edge states in Lindbladian dynamics, however, remain unexplored. A central aspect in closing this knowledge gap is that dynamics is governed not only by the localization of eigenstates, but also by their complex spectrum. Extremal edge modes and chiral damping have been observed in different classes of systems, the former in examples with spectral line-gap topologies~\cite{Hegde23}, and the latter with spectral point-gap topologies~\cite{Yang_2023}. In this work, we show that chiral damping and edge-selective extremal damping can coexist in open quantum systems due to a non-trivial interplay of (i) spectral topology in the bulk giving rise to an NHSE, (ii) band topology in the bulk giving rise to topological edge states, and (iii) spectral topology at the edge giving rise to a damping gap and minimal damping of the edge states. In addition, we explain how the time-evolution of particle densities provides a convenient probe of these different topologies.

We focus on the example of a Hofstadter model~\cite{hofstadter1976energy} coupled to an environment inducing gain and loss along the $y$-bonds, see Fig.~\ref{fig:model}(a). Our main results are as follows. (A) Since bulk sites have two attached $y$-bonds, while edge sites have only one, bond dissipation generically gives rise to weaker dissipation at the edges than in the bulk. Physically, this implies that the states  localized at the edges tend to have the lowest decay. (B) At the same time, bond dissipation is well-known to induce an NHSE~\cite{Yao_2018,Zhang_2020,Bergholtz_2021}, and we observe the corresponding chiral damping wavefront in the bulk~\cite{Song2019,Yang2022}. The edge states, however, experience a competition of topological and skin localizations, both exponential. For sufficiently small dissipation, the edge states resist skin localization~\cite{Zhu2021,Cheng2022,Zhu2022,Wang2022Morph}. We then find that particle localization at the edges persists even at long times. (C) This physics can be detected via the spatially resolved particle density in the system, as schematically illustrated in Fig.~\ref{fig:model}(b), explaining the main features of a numerical simulation shown in Fig.~\ref{fig:model}(c). (D) We connect our concrete results to a general discussion of the interplay of spectral and band topologies, explaining how our findings pertain to dynamics in open quantum systems more broadly. \\

{\it Model.} We consider spinless electrons on a two-dimensional square lattice subject to a uniform magnetic field, see Fig.~\ref{fig:model}(a). With the gauge choice $\mathbf{A} = B (-y,0)$, applying Peierls substitution~\cite{hofstadter1976energy}, and assuming cylindrical boundary conditions (periodic in $x$, open in $y$), the Hamiltonian  takes the form

\begin{equation} \label{eq:ham}
\begin{split}
H = \sum_{k_x,j_y}\Bigl( 2t_x \cos{(k_x - 2 \pi j_y \phi/\phi_0)} c_{k_x,j_y}^\dagger c_{k_x,j_y}^\pdag \\+  t_y^R c_{k_x,j_y+1}^\dagger c_{k_x,j_y}^\pdag+ t_y^L c_{k_x,j_y}^\dagger c_{k_x,j_y+1}^\pdag\Bigr)\:,
\end{split}
\end{equation}
where $c_{k_x,j_y}^\dagger$ are fermionic creation operators for an electron with $x$-momentum $k_x$ and site number $j_y$ along $y$. The hopping amplitudes are $t_x, t_y^{R/L}$, the magnetic flux per unit cell is $\phi$, and $\phi_0 = h/e$ is the flux quantum. An isolated Hofstadter system has $t_y^R=t_y^L=t_y$, and is thus a Hermitian Hamiltonian. Using greek indices as composites for $k_x$ and $j_y$, the Hamiltonian defines a matrix $h$ via $H = \sum_{\alpha\beta} c_{\alpha}^\dagger\,h_{\alpha\beta} c_{\beta}^\pdag$.

The environment is introduced in the language of a Lindblad Master equation for the system's density matrix $\rho$~\cite{lindblad1976generators,breuer2002theory},
$
\frac{d\rho}{d t} = -i [H, \rho] + \sum_\alpha \left( L_\alpha \rho L_\alpha^\dagger - \frac{1}{2} \{ L_\alpha^\dagger L_\alpha, \rho \} \right),
$
where dissipation is described by the jump operators
\begin{subequations}
\begin{align}
L^l_{k_x,j_y} &= \sqrt{\gamma^l} \,(c_{k_x,j_y}^\pdag - i c_{k_x,j_y+1}^\pdag), \quad\\
L^g_{k_x,j_y} &= \sqrt{\gamma^g} \,(c_{k_x,j_y}^\dagger + i c_{k_x,j_y+1}^\dagger).
\end{align}\label{eq:jump_operators}
\end{subequations}
Here, $\gamma^l$ ($\gamma^g$) represents the dissipation strength for loss (gain). 
Our main observable of interest is the covariance matrix $C$ with matrix elements $C_{\alpha\beta}(t) = \text{Tr}(c_{\alpha}^\dagger c_{\beta}^\pdag \rho(t))$. Its time evolution is governed by~\cite{Song2019}

\begin{equation}
\frac{dC(t)}{dt} = X C(t) + C(t) X^\dagger + 2 M^g,
\label{eq:lyapunov}
\end{equation}
where $M^g$, $M^l$, and the damping matrix $X$ are defined as $M^g_{\alpha\beta}=\sum_\nu D^{g\:*}_{\nu\alpha} D^g_{\nu\beta}$,  $M^l_{\alpha\beta}=\sum_\nu D^{l\:*}_{\nu\alpha} D^l_{\nu\beta}$, and $ X=i h^T-({M^l}^T+M^g)$ after rewriting $L^g_\nu = \sum_\alpha D^g_{\nu\alpha} c_\alpha^\dagger, \quad L^l_\nu = \sum_\alpha D^l_{\nu \alpha} c_\alpha^\pdag$. The solution of Eq.~\eqref{eq:lyapunov} can be split into a constant steady-state value $C_{\rm ss}$ and the convergence towards the steady state as ${C}(t) = C_{ss} + \tilde{C}(t)$. The time evolution of $\tilde{C}(t)$ can be cast into an analytic expression in terms of the eigensystem of $X$,

\begin{equation} \label{eq:covariance_evolution}
\tilde{C}(t) = \sum_{m,m'} e^{(\lambda_m + \lambda^*_{m'})t} | \psi_{R,m} \rangle \langle \psi_{L,m} | \tilde{C}(0) | \psi_{L,m'} \rangle \langle \psi_{R,m'} |,
\end{equation}
where $\lambda_m$  ($\psi_{R/L,m}$) denotes the $m$-th eigenvalue (right/left eigenvector) of the damping matrix $X$~\cite{Song2019}. \\

{\it Interplay of spectral and band topologies.} Our model is one example of a system exhibiting both spectral and band topologies. In our case, both topologies are hosted by the damping matrix $X$, but similar conclusions can also be drawn for non-Hermitian Hamiltonians and other non-Hermitian matrices.

Consider therefore a Hermitian matrix $\mathcal{M}$ representing some topological system with open boundary conditions along $y$ (such as our damping matrix for $\gamma^l=\gamma^g=0$, or a usual Hamiltonian). We furthermore assume the system to be coupled to some environment giving rise to  non-reciprocal nearest-neighbor hoppings $t_y \pm \gamma$, which entails an NHSE along $y$. In the simplest cases, the non-Hermiticity corresponding to this non-reciprocity can be removed by virtue of a so-called imaginary gauge transformation (IGT), $\mathcal{M} (t_y,\gamma ) =S^{-1} \mathcal{M}(\tilde{t}_y,0 ) S$, where $S$ implements the IGT and $\tilde{t}_y = \sqrt{t_y^2- \gamma^2}$~\cite{Hatano_Nelson_1996,Yao_2018,Yokomizo_2019}.  This transformation preserves the spectrum but scales the eigenstates with an exponential factor $e^{y/\xi_S}$, where $\xi_S = \left( \frac{1}{2} \log \frac{t_y-\gamma}{t_y+\gamma} \right)^{-1}$, thus removing the NHSE.

Since the system is topological by assumption, bulk-boundary correspondence implies that the Hermitian matrix $\mathcal{M}(\tilde{t}_y,0 )$  generically has exponentially localized, topologically protected edge states. A reverse IGT then shows that while all extended bulk states of $\mathcal{M} (\tilde{t}_y,0)$ correspond to skin-localized states of $\mathcal{M}({t}_y,\gamma)$, the same need not be true for the edge states. Those instead, experience a competition between topological and skin localization: one edge mode is tightly localized, $e^{-y/\xi_S}e^{-y/\xi_R}$, while the other is stretched out, $e^{-y/\xi_S}e^{-(L-y)/\xi_L}$, where $\xi_{R/L}$ are the topological localization lengths. If $\xi_S \gg \xi_{R,L}$, i.e.~for sufficiently weak non-reciprocity, both edge modes remain localized at their respective edges. Bulk band topology (giving rise to topological edge states) and bulk spectral topology (yielding a bulk NHSE) thus co-govern the spatial profile of eigenstates~\cite{Zhu2021,Cheng2022,Zhu2022,Wang2022Morph}.

The central interest of the present work is how the interplay of spectral and band topologies affects dynamics. To boost the impact of the rebellious topological edge states onto dynamics, the damping rates for bulk and edge states should differ significantly. In an initially Hermitian topological system, a corresponding gap in damping rates(i.e real part of $X$-spectrum) can be induced by adding dissipation only to the edges.Technically, this leads to a spectral point gap for the topological edge states~\cite{Nakamura2023}. In open quantum systems, effective edge-only dissipation can readily be created by bond dissipators, since edge sites have less bonds attached than bulk sites. It is now important to recall that (reverse) IGTs not only preserve the spectrum of $\mathcal{M}$, but also its diagonal elements. Adding edge-only dissipation to either $\mathcal{M} (\tilde{t}_y,0)$ or $\mathcal{M} (t_y,\gamma )$ consequently results in a damping gap for (topological) edge states, but not for any other states. In particular, skin-localized bulk states of $\mathcal{M} (t_y,\gamma )$ are essentially unaffected. \\

{\it Deconstructing damping effects in the spectrum of a dissipative Hofstadter model.} 
The non-Hermitian Hofstadter model and related Aubry-Andr\'e-Harper model have been studied before in the context of suppression of skin-effect and chiral damping as well as localization-delocalization transitions~\cite{Purkayastha_2018,Longhi2019, Zhang2020,Zeng2020,Longhi2021,Liu2021,Tang2021,He2022,Mallick2023,Zhou2024,Padhi2024}. Here, we are interested in how the interplay of spectral and band topologies affects dynamics, especially at intermediate magnetic fields (we comment on weak fields later). For the Hofstadter model with bond dissipators defined in Eqs.~\eqref{eq:ham} and \eqref{eq:jump_operators}, the damping matrix reads

\begin{equation} \label{eq:ham_edge}
X = i \,H_{\text{nH-Hof}} - 2\gamma\, \mathds{1} + \gamma \,I_{\text{edges}},
\end{equation}
where  $\gamma = \gamma^g + \gamma^l$ and $H_{\text{nH-Hof}} = \left.H\right|_{t_y^R=t_y+\gamma,~t_y^L=t_y-\gamma}$ is a non-Hermitian Hofstadter model defined by Eq.~\eqref{eq:ham} with non-reciprocal hoppings. The term $\sim\mathds{1}$ merely shifts the entire spectrum of $X$ by $2\gamma$, but leaves all eigenstates unaffected. In contrast, $I_{\text{edges}}$ is a diagonal matrix with zeros everywhere except for unit entries at the top left and bottom right (corresponding to the edge sites). This term physically encodes the fact that edge sites have fewer attached bonds than bulk sites, and are consequently less affected by bond dissipators.

\begin{figure}[t]
    \centering
    \includegraphics[width=0.4\textwidth]{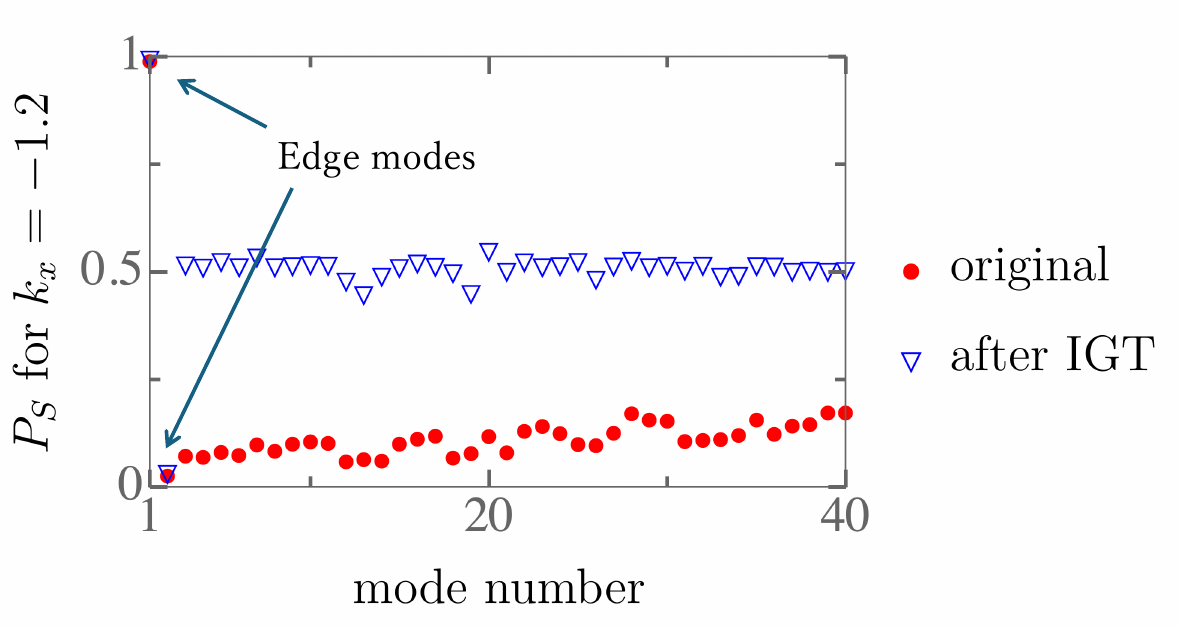}
    \caption{\label{fig:PS}\textbf{Deconstructing band topology and spectral topology.} Static polarization $P_S$ for the original eigenstates of the damping matrix $X$, and after removing the NHSE using an IGT. Here, $\gamma^l=0.2$, $\gamma^g=0$, $\phi/\phi_0=1/4$, $t_x=t_y=1$, and $L_y=40$.}
\end{figure}

We first analyze the spatial localization of the eigenstates of $X$, captured by the spectral polarization 

\begin{equation}
    P_S(m,k_x)= \frac{1}{L_y}\sum_{j_y=1}^{L_y} j_y\: |\psi_{R,m}(k_x,j_y)| ^2.
\end{equation}
Here, we used translation invariance along $x$ to decompose the problem into decoupled $k_x$-sectors. In our numerical implementations, we normalize the right eigenstates as $\sum_{j_y}|\psi_{R,m}(j_y,k_x)| ^2=1$. Fig.~\ref{fig:PS} shows the spectral polarization in a parameter regime supporting topological edge states. To allow a convenient identification of bulk and edge states, we depict $P_S$ in the original model (red dots), and also after removing the NHSE using an IGT (blue open triangles). As heralded by a value of $P_S\approx0$, the NHSE localizes most states at one edge. In striking contrast, the topological edge states (modes 1 and 2 in Fig.~\ref{fig:PS}) remain well-localized at their respective edges despite the NHSE.

Next, we turn to the dynamics of electronic densities. The momentum-resolved time-dependence is described by the dynamical polarization, defined as

\begin{figure*}[t]
    \centering
    \includegraphics[width=0.99\textwidth]{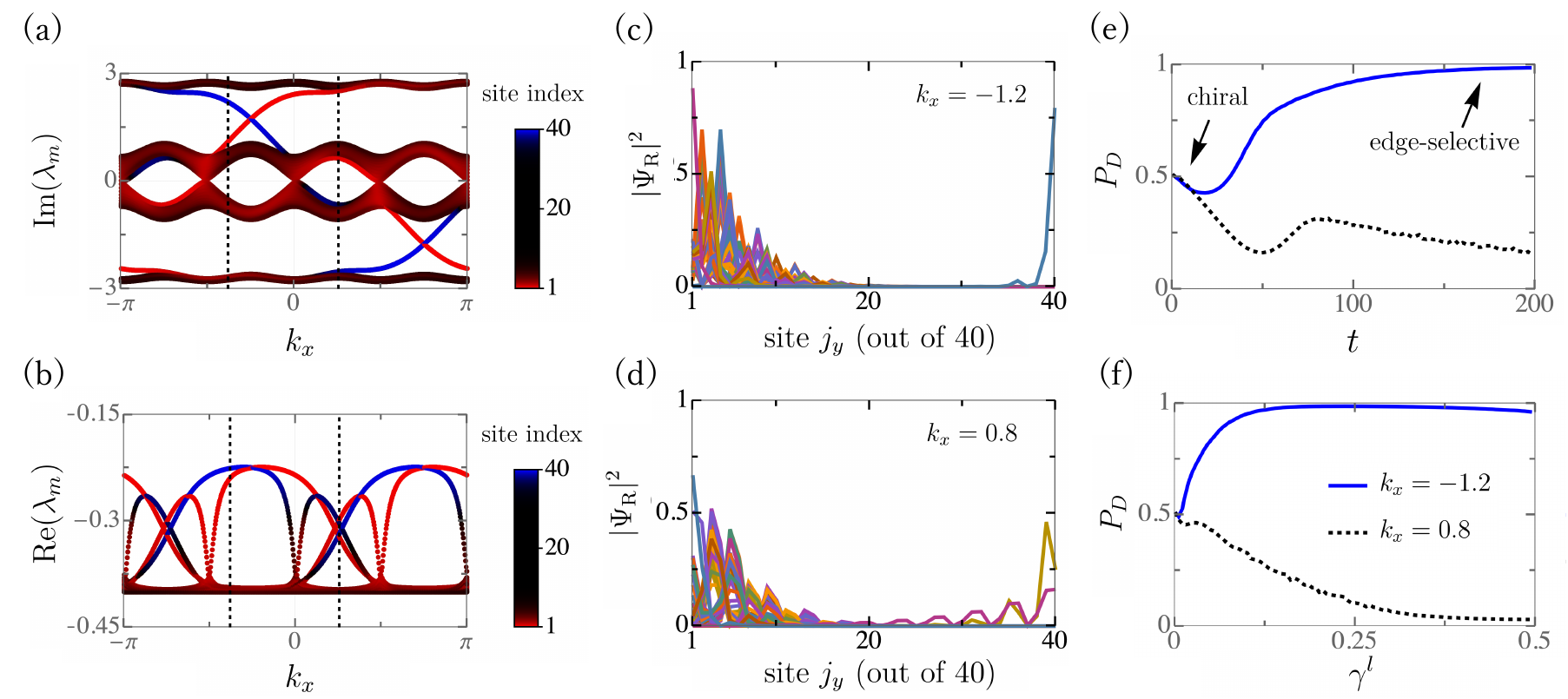}
    \caption{\label{fig:dynamics} \textbf{Dynamics: coexistence of topological edge modes and skin-localized bulk modes.} (a,b) Imaginary and real parts of $\lambda_m$ (eigenvalues of $X$). Colors indicate the site index expectation value of the corresponding right eigenstates, vertical lines mark $k_x=-1.2$ and $k_x=0.8$. (c,d) Spatial profile of the eigenstates for $k_x=-1.2$ and $k_x=0.8$. (e) Dynamical polarization as a function of time. (f) Dynamical polarization as a function of damping at fixed time $t=200$. Here, $\gamma^l=0.2$ except for (f), $\gamma^g=0$, $\phi/\phi_0=1/4$, $t_x=t_y=1$, and $L_y=40$.}
\end{figure*}

\begin{equation}
    P_D(k_x,t) = \frac{1}{L_y}\frac{\sum_{j_y=1}^{L_y} \:j_y \:n(k_x,j_y,t)}{\sum_{j_y=1}^{L_y} \:n(k_x,j_y,t)}.
\end{equation}

Here, the time-dependent local density $n(k_x,j_y,t)$ is simply the diagonal element of $C(t)$ with $\alpha=\beta=(k_x,j_y)$. Since Eq.~\eqref{eq:ham_edge} contains a term corresponding to edge-only dissipation, our general discussion implies that the edge states exhibit a damping gap. This is shown in the $k_x$-resolved spectrum of the damping matrix in Fig.~\ref{fig:dynamics}(a)-(b). The imaginary part of the spectrum reflects the band-structure of the Hofstadter model. Topological edge states crossing the band gaps are clearly visible in Fig.~\ref{fig:dynamics}(a) around $k_x\approx -0.8$ and $k_x\approx 2.4$. The real part of the spectrum depicted in Fig.~\ref{fig:dynamics}(b) shows that edge-only dissipation induces a damping gap for these edge states. A simple perturbative argument indicates that the edge-state damping gap is of order $\gamma$, in agreement with our numerical findings. 

Due to this damping gap, the edge states govern the long-time dynamics. We show this using two representative momenta: well-defined topological edge states exist for $k_x=-1.2$, while they are not well-separated from bulk states for $k_x=0.8$, as visible from Fig.~\ref{fig:dynamics}(a)-(b), and further illustrated in Fig.~\ref{fig:dynamics}(c)-(d) showing the spatial profile of the right eigenstates of $X$. Note that for a fixed $k_x$, different edge states in general have different distances to the bulk states. This results in different localization lengths and different damping gaps. The long-time dynamics for fixed $k_x$ is governed by the single edge state with the lowest damping. Fig.~\ref{fig:dynamics}(e) presents the time-dependence of $P_D$. We find that the chiral damping rooted in the NHSE of $X$ leads to an initial linear evolution (here: decrease) of the dynamical polarization. In contrast, the long-time dynamics depends on whether well-defined edge states with a damping gap are present or not: if present, such edge states induce edge-selective extremal damping. To assess the robustness of this effect with varying dissipation, we plot the polarization at the late time $t=200$ as a function of the damping $\gamma$ in Fig.~\ref{fig:dynamics}(f). For $k_x = -1.2$, we find that $P_D$ first settles to 1, heralding edge-selective extremal damping, but decreases as the NHSE becomes stronger at larger $\gamma$, which reduces the localization of the edge state. For $k_x = 0.8$, on the contrary, edge-selective extremal damping is absent due to the absence of well-defined damping gap for the edge states. This demonstrates that well-defined topological edge states are indeed resilient against skin localization, and that they facilitate edge-selective extremal damping due to their damping gap. Particles will consequently predominantly remain at the edges at late times.

The momentum-resolved analysis of Fig.~\ref{fig:dynamics} can be generalized to more accessible observables. For the setting of Fig.~\ref{fig:model}, the momentum-summed local densities $n(j_y,t) = \sum_{k_x} n(k_x,j_y,t)$, summed over 40 equidistant $k_x$-values, exhibit two different regimes. After preparing the system in a fully occupied state, the short-time dynamics is governed by an emptying of the bulk states via a chiral damping wavefront, see Fig.~\ref{fig:model}(c). The topological edge states, however, govern the long-time dynamics due to their damping gap: only particles at the edges remain in the system at long times. This illustrates that local particle densities are a convenient probe for the coexistence of spectral topology and band topology. \\

\begin{figure} [t]
    \centering
    \includegraphics[width=0.47\textwidth]{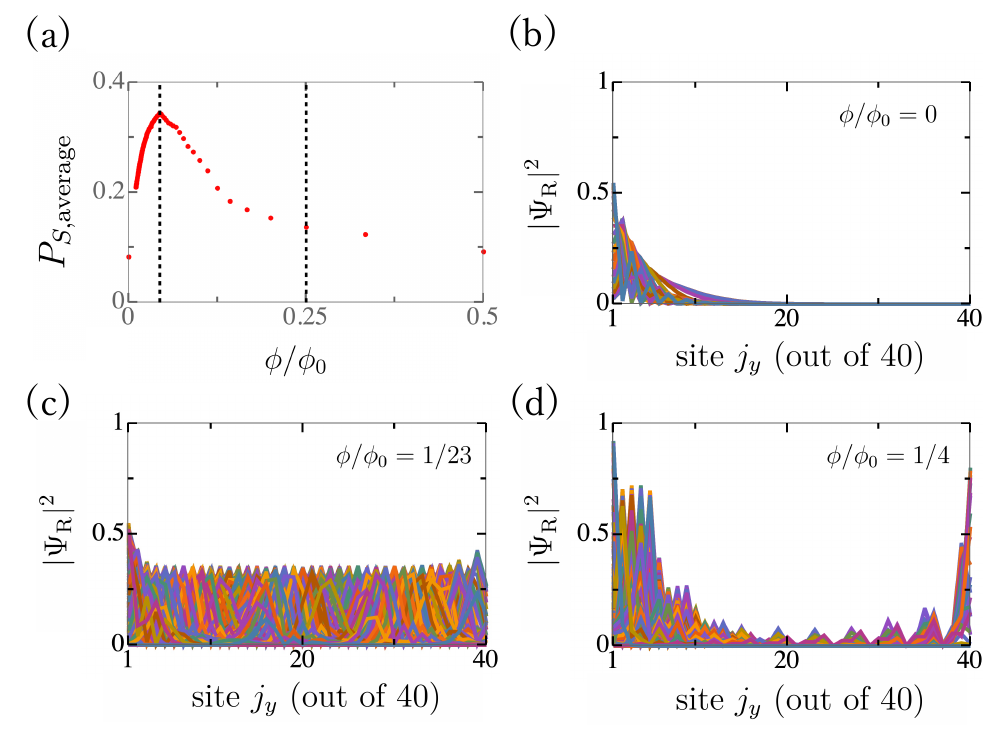}
    \caption{\label{fig:static} \textbf{Magnetic suppression and recovery of skin localization.} (a) Static polarization averaged over $k_x$ at fluxes $\phi/\phi_0 = 1/\nu$ with $\nu=2,\ldots,100$ and $\phi=0$. Dashed lines indicate $\phi/\phi_0=1/23, 1/4$. (b,c,d) Spatial profile of the right eigenstates of $X$ for all momenta in the average. The fluxes are $\phi/\phi_0=0$ in (a), $\phi/\phi_0=1/23$ in (b), and $\phi/\phi_0=1/4$ in (c). We use $\gamma^l=0.2$, $\gamma^g=0$, $t_x=t_y=1$, and $L_y=40$.
    }
\end{figure}

{\it Magnetic field-dependence of skin localization.} 
The NHSE is known to be suppressed by weak magnetic fields because spatially localized Landau level states differ dramatically from extended Bloch states~\cite{Lu_2021}, which also affects chiral damping~\cite{He2022}. In agreement with this result, we observe that $P_{S,\text{average}}$, the spectral polarization averaged over 40 equidistant $k_x$-momenta, rapidly increases as the magnetic field is switched on, see Fig.~\ref{fig:static}(a). When the flux per plaquette reaches intermediate values, however, the localization profile of eigenstates changes, see Fig.~\ref{fig:static}(b)-(d). We find that an extensive skin effect is recovered at intermediate fluxes, rendering chiral damping observable again.\\

{\it Conclusions.} We find that the dynamics of particle densities in open quantum systems enables a direct visualization of coexisting spectral and band topologies. In particular, we show that the non-Hermitian skin effect, a paradigmatic example of spectral topology, can dominate the short-time dynamics, while band topology can protect topological edge states governing the long-time dynamics. The measurable consequence thereof is a combination of chiral damping at short times and edge-selective extremal damping at long times. We explain that the decoupling of chiral damping and edge-selective extremal damping in time is due to their separate origins (non-reciprocal hopping vs.~edge-specific damping). While we discuss these results with the example of a dissipative Hofstadter model, in which the individual analysis of spectral and band topologies is particularly illuminating, our findings provide insights into the interplay of spectral and band topologies in the dynamics of open topological quantum systems broadly.\\

{\it Acknowledgments: } R.S. is supported by the Prime Minister's Research Fellowship (PMRF). A.N. acknowledges support from the DST MATRICS grant (MTR/2023/000021). S.H and T.M. acknowledge funding by the Deutsche Forschungsgemeinschaft (DFG) via the Emmy Noether Programme (Quantum Design grant, ME4844/1, project-id 327807255), project A04 of the Collaborative Research Center SFB 1143 (project-id 247310070), and the Cluster of Excellence on Complexity and Topology in Quantum Matter ct.qmat (EXC 2147, project-id 390858490).

$^*$ These authors contributed equally.
\bibliography{main.bib}

\end{document}